\documentclass[pdflatex,twocolumn,a4paper,oneside,sn-mathphys-ay]{sn-jnl}

\usepackage{graphicx}%
\usepackage{multirow}%
\usepackage{amsmath,amssymb,amsfonts}%
\usepackage{amsthm}%
\usepackage{mathrsfs}%
\usepackage[title]{appendix}%
\usepackage{xcolor}%
\usepackage{textcomp}%
\usepackage{manyfoot}%
\usepackage{booktabs}%
\usepackage{algorithm}%
\usepackage{algorithmicx}%
\usepackage{algpseudocode}%
\usepackage{listings}%

\let\orcidlogo\relax 
\usepackage{orcidlink}
\usepackage{braket}
\usepackage{bbm}
\usepackage{url}

\usepackage{geometry}
\usepackage[acronym,nonumberlist]{glossaries}

\newacronym{ml}{ML}{Machine Learning}
\newacronym{qc}{QC}{Quantum Computing}
\newacronym{vrp}{VRP}{Vehicle Routing Problem}
\newacronym{cvrp}{CVRP}{Capacitated Vehicle Routing Problem}
\newacronym{vrptw}{VRPTW}{Vehicle Routing Problem with Time Window}
\newacronym{cvrptw}{CVRPTW}{Capacitated Vehicle Routing Problem with Time Window}
\newacronym{mdvrp}{MDVRP}{Multi-Depot Vehicle Routing Problem}
\newacronym{tn}{TN}{Tensor Network}
\newacronym{qubo}{QUBO}{Quadratic Unconstrained Binary Optimization}
\newacronym{qaoa}{QAOA}{Quantum Approximate Optimization Algorithm}
\newacronym{qa}{QA}{Quantum Annealing}
\newacronym{mps}{MPS}{Matrix Product State}
\newacronym{qml}{QML}{Quantum Machine Learning}
\newacronym{nisq}{NISQ}{Noise Intermediate Scale Quantum}
\newacronym{gd}{GD}{Gradient Descent}
\newacronym{bqp}{BQP}{Bounded-Error Quantum Polynomial}

\theoremstyle{thmstyleone}%

\theoremstyle{thmstyletwo}%

\theoremstyle{thmstylethree}%

\glsdisablehyper


\begin{document}

\title{Quantum Vehicle Routing Problem: a smart initialization strategy for ill-conditioned optimization landscapes}

\author[1]{\orcidlink{0000-0003-3840-3609} \fnm{Emanuele} \sur{Casciaro}} \email{emanuele.casciaro@unifi.it}

\author[2]{\fnm{Mario} \sur{De Simone}} \email{mario.desimone@edu.unifi.it}

\author*[1]{\orcidlink{0000-0001-6024-1752} \fnm{Stefano} \sur{Martina}} \email{stefano.martina@unifi.it}

\author[3]{\orcidlink{0009-0008-6980-7095} \fnm{Fabio} \sur{Mascherpa}} \email{fabio.mascherpa@eni.com}

\author[3]{\orcidlink{0000-0002-7473-5945}~\fnm{Alfonso} \sur{Amendola}} \email{alfonso.amendola@eni.com}

\author[1]{\orcidlink{0000-0002-8366-4296} \fnm{Filippo} \sur{Caruso}} \email{filippo.caruso@unifi.it}

\affil[1]{\orgdiv{Department of Physics and Astrophysics}, \orgname{University of Florence}, \orgaddress{\street{Via Sansone, 1}, \city{Sesto Fiorentino}, \postcode{50019}, \country{Italy}}}

\affil[2]{\orgdiv{Department of Information Engineering}, \orgname{University of Florence}, \orgaddress{\street{Via di S. Marta, 3}, \city{Florence}, \postcode{50139}, \country{Italy}}}

\affil[3]{\orgdiv{DICOX/C – High Performance Computing Center of Excellence, DIT – Digital \& Information Technology}, \orgname{Eni S.p.A.}, \orgaddress{\street{ Via Emilia 1}, \city{San Donato Milanese}, \postcode{20097}, \country{Italy}}}

\abstract{
    With the expansion of complex transport infrastructure, new paths for linking different destinations have emerged, allowing the definition of multiple routes. The introduction of new roads may also lead to disruptions in traffic flow.
    This problem can be generalized into a special case of \gls{vrp}, which aims to find the optimal route for a fleet of vehicles within a network of destinations.
    Finding solutions to this problem requires an exponential number of operations, hindering the ability of exact solvers to deal with large instances,
    which must therefore be tackled using heuristics, often leading to suboptimal results.
    Quantum-based algorithms may allow us to find exact solutions in polynomial time with high probability, offering a potential future alternative to classical solvers and in this work, we focus on the \gls{qaoa}. Our novel contributions are a parameter-initialization strategy that improves the convergence probability and the integration of the subtour constraints directly into the optimization without using additional qubits by employing an approximation of exponential equality constraints. We believe that our results pave the way for effectively solving instances of \gls{vrp} using quantum-based solutions, leveraging the space-efficient problem encoding and our initialization strategy, which limits the effect of barren-plateau under random parameter initialization.
    }


\maketitle

\section{Introduction}
The resolution of high-complexity combinatorial problems still represents an open branch of mathematics and computing, with continual research for better solving algorithms capable of producing higher quality solutions in less time and ideally providing mathematically-provable exact solutions to the problems of interest. With combinatorial optimization, it is possible to solve a wide range of problems, highlighting its relevance in the field of logistics, where it can be used to find the optimal way of handling available resources~\citep{boria_survey_2011}, to optimize industrial processes~\citep{crama_combinatorial_1997}, and to improve healthcare systems~\citep{abdalkareem_healthcare_2021}, aiding in the definition of optimal schedules aimed at minimizing management costs. Another relevant field of application is the management of vehicle traffic routes, i.e., a topic that has become more relevant with the spread of commercial vehicles, making traffic routing a crucial component of daily life with a significant impact on delivery systems and traffic management. As cities expand and new routes are created with the goal of improving traffic flow, the resulting road network can become less efficient without proper traffic routing~\citep{braess_paradox_2005}. The efficient routing of vehicles is also important to effectively reduce the cost of deliveries by grouping one or more expeditions whenever possible. This task results in a more effective use of available time and fuel, with subsequent economic and logistical advantages.
This traffic routing problem is globally recognized by the name \gls{vrp} and is a well-known instance of NP-Hard~\citep{Karp1972} optimization. 

The \gls{vrp} can be further extended to model more complex scenario, by adding additional constraints, including (but not limited to): \gls{cvrp} where the maximum volume / weight that each vehicle can carry is limited and customers request specific amounts of capacity, \gls{vrptw} where each customer requests that a vehicle perform the delivery within a desired time frame, \gls{mdvrp} where each vehicle is assigned to one of the multiple depots available within the network, or combinations of them such as \gls{cvrptw}. However, it is beyond the scope of this article to analyze these variations in detail, and we refer the reader to~\citet{kumar_survey_2012} to delve further into the subject. This family of problems is well known to be \textit{NP-hard}~\citep{arora2006computational, Karp1972}, which poses a challenge to current solving techniques as they require an exponential number of operations to solve them.

In the current state of the art, the \gls{vrp} can be solved using either exact or approximate methods. The former are generally based on branch-and-bound algorithms~\citep{laporte_branch_1983, carpaneto_branch_1989} that generate new solutions until a global optimum is found. These solvers are often part of general-purpose optimization suites, such as Gurobi or IBM CPLEX.
They offer flexible frameworks to solve instances exactly, at the cost of limiting the instance size. Alternatively, approximate methods rely on evolutionary algorithms and heuristics~\citep{liu_heuristics_2023} with the goal of generating solutions that are close to the true optimum. A plethora of solvers exists to deal with different variations of the problem, using ad hoc implementations designed to find solutions to particular \gls{vrp} subproblems in a short amount of time. These architectures, despite requiring approximations, enable solving larger problems and are the standard approach in commercial settings.

With the advent of \gls{qc}~\citep{preskill2018quantum, cerezo2021variational}, new approaches to solve combinatorial problems have emerged. In particular, quantum computers are naturally suited to solve \gls{qubo} problems~\citep{kim_quantum_2025}, either via \gls{qa}~\citep{rajak_quantum_2023} (used in analog \gls{qc}) or through its digital counterpart \gls{qaoa}~\citep{farhi_quantum_2014}, which is the focus of this article. The computational power of quantum devices is captured by the complexity class \gls{bqp}~\citep{qcomplexities}. Although NP-hard problems are not believed to be efficiently solvable even by quantum computers, quantum heuristics such as \gls{qa} and \gls{qaoa} explore the solution space through fundamentally different mechanisms, namely superposition, entanglement and interference. This may yield practical speed-ups or higher-quality solutions on specific problem classes. 
To exploit this potential, a problem must first be expressed as an Ising Hamiltonian: both \gls{qa} and \gls{qaoa} rely on the one-to-one correspondence between a \gls{qubo} formulation of the problem and a Hamiltonian whose ground state encodes the optimal solution. The quantum device is then used to prepare states with a large overlap with this ground state. The bottleneck of this approach lies in the conversion of a generic combinatorial problem into the \gls{qubo} formalism. This step, despite being always possible, may require--in case of inequality constraints--additional variables and generally require the introduction in the model of hyper-parameters that cannot be directly optimized. This overhead in the problem, given the inherit nature of combinatorial problem, grows quickly in the size of the problem, resulting in a \gls{qubo} formulation with an exponential number of terms.
Beyond the formulation, the training of \gls{qaoa} is itself challenging. The classical optimization of the variational parameters is hindered by local minima and barren plateaus~\citep{mcclean_barren_2018}, i.e., regions where the gradient vanishes exponentially with system size. Several strategies have been proposed to mitigate this, including layer-wise interpolation of parameters~\citep{zhou2020quantum}, warm-starting from classical relaxations~\citep{egger_warm-starting_2021}, and parameter transfer between instances. Many of these, however, depend on problem-specific information or on solving auxiliary problems.

The \gls{vrp} has already been explored in the context of quantum-assisted optimization. The literature on this problem, and more generally quantum combinatorial optimization, can be grouped into two categories: \gls{qa} solvers~\citep{morita_mathematical_2008} and \gls{qaoa} solvers~\citep{farhi_quantum_2014}. Even though both share the exact problem formulation requirements, the former is suitable for analog quantum computing platforms, like the DWave quantum annealers, which apply a continuous evolution to the system. The latter, instead, uses the circuit-based computational model to carry out the optimization and is more suitable for digital quantum devices such as IBM machines. The \gls{qa} based approach can rely on a broader qubit count, allowing the resolution of larger problems. However, the execution time must be long enough to respect the \textit{adiabatic theorem}. The large registry size of the DWave 2000 device allowed \citet{irie_quantum_2019} to solve \gls{cvrptw} up to 7 nodes, achieving $80\%$ of generated valid solutions.
Another independent work solved \gls{cvrptw} using a different encoding of the problem, managing to tackle configurations up to 12 nodes.
The larger qubit count has also been combined with classical techniques; for instance, \citet{borowski_new_2020} decompose the \gls{cvrp} through clustering algorithms into several instances of the Traveling Salesman Problem~\citep{laporte_traveling_1992}, solved independently using DWave Leap, achieving an optimality gap~\citep{nemhauser_general_1988} of $25.13\%$ over a problem instance with $101$ nodes and $8$ vehicles.

A similar approach is used by \citet{feld_hybrid_2019}, in which the \gls{cvrp} instance is decomposed into two distinct optimization problems. The first problem, solved classically, is related to the clustering of the graph and is associated with an instance of the Knapsack Problem~\citep{korte_knapsack_2008}; after this step, a quantum annealer is used to generate the routes on the clustered graph.

The problem has also been approached with \gls{qaoa} solutions, with promising early results. In \citet{azfar_quantum-assisted_2026}, the \gls{vrp} for 3 nodes is successfully solved using IBM Quantum System One, showing that the optimization can be performed even with \gls{nisq} devices.
With \gls{qaoa}, it is also possible to solve constrained problems; a good example is given by \citet{xie_feasibility-preserved_2024}, in which the authors use an additional ancilla register together with a mixer that preserves feasibility for solving \gls{cvrp} with 5 nodes.
A different approach is adopted by \citet{huang_solving_2025}, which, inspired by classical exact solvers, employs \gls{qaoa} alongside the column generation technique, solving \gls{cvrp} up to 7 nodes.

In this work, we further explore the application of \gls{qaoa} to the \gls{vrp}, contributing mainly in the following aspects. First, we propose an initialization strategy for \gls{qaoa}, aimed at mitigating optimization in ill-conditioned landscapes often present in quantum optimization routines. Then, building on \citet{montanez-barrera_unbalanced_2024}, we introduce an unbalanced penalization term to remove the exponential parameter overhead that is usually required to define the \gls{vrp} in the \gls{qubo} formalism. To the best of our knowledge, both the proposed initialization strategy and the definition of the subtour \gls{qubo} term represent a novelty in the context of \gls{qaoa} and \gls{vrp} resolution methods. 
The remainder of the article is organized as follows: Section~\ref{sec:vrp} provides a comprehensive introduction to \gls{vrp} and its formalization as a \gls{qubo} problem and the implementation using \gls{qaoa}. In Section~\ref{sec:init} we introduce our proposed novel initialization strategy for \gls{qaoa}. In Section~\ref{sec:subtour}, we propose an encoding capable of including the subtour constraint, obtaining a correct solution with high probability; to the best of our knowledge, no other end-to-end \gls{qaoa} routine in the literature manages to integrate and solve the full \gls{vrp} problem for a generic instance without an overhead in circuit size.
A summary of the results is reported and discussed in Section~\ref{sec:conclusion}.

\section{Vehicle Routing Problem}\label{sec:vrp}
\begin{figure*}[h!]
    \centering
    \includegraphics[width=0.8\linewidth]{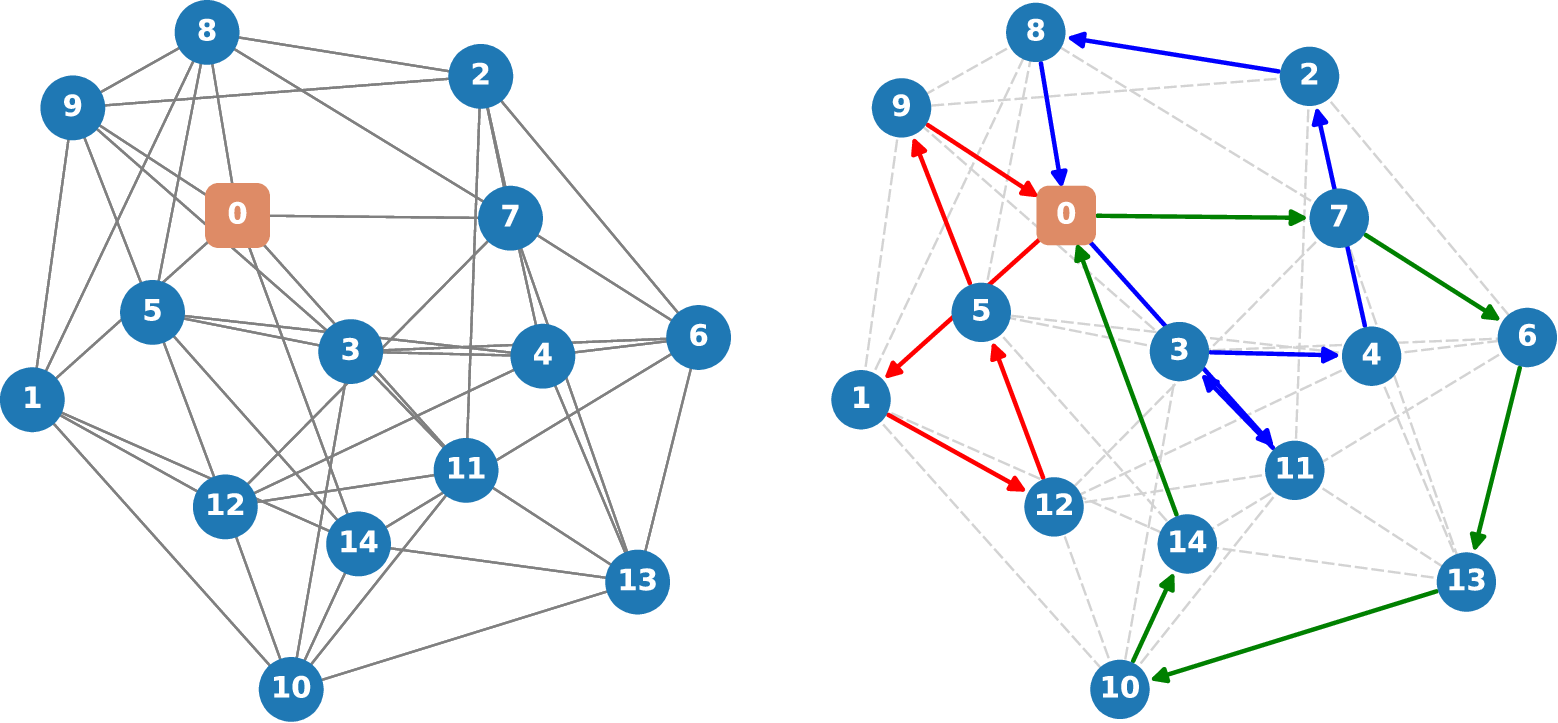}
    \caption{Example of VRP problem with $N=15$ nodes and $K=3$ vehicles. Given a base graph (left) and a cost associated with the travel from one node to the other, the goal is to determine the optimal path of each of the $K$ vehicles (colored arrows on the right) such that every node (blue circles) except the depot (orange square) is reached exactly one time and all the vehicles start and end in the depot.}
    \label{fig:example}
\end{figure*}

\begin{figure*}[h!]
    \centering
        \includegraphics[width=0.6\linewidth]{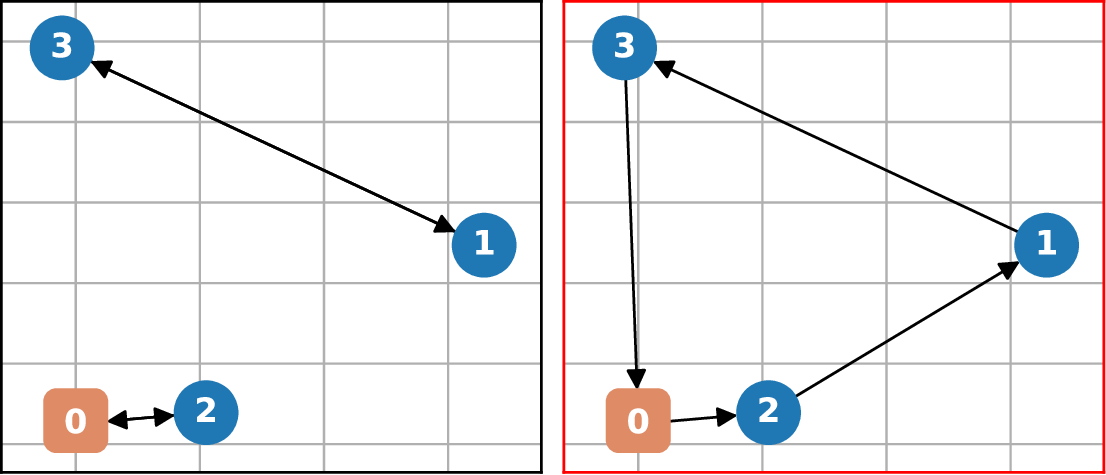}
        \caption{Two example solutions for the problem $VRP(4, 1, W)$ with $4$ destinations and one vehicle. The solution obtained without encoding the subtour constraint (left, black frame) presents disconnected loops and result in a configuration admissible only if we have multiple depots. Including the subtour constraint allows to find the correct solution (right, red frame).}
    \label{fig:subtour-example}
\end{figure*}

The \gls{vrp}, depicted in Figure~\ref{fig:example}, is formalized as a set of $K$ vehicles that move on a graph $G=(V,E)$ of $N=|V|$ vertices where one of them (conventionally the one with the first index) is a special depot node and all are connected by weighted edges $E$.
The goal of \gls{vrp} is to determine the optimal route for each vehicle such that every node is visited exactly once, with the exception of the depot, which acts as the starting and final point for every element of the fleet. 
We can then identify a single \gls{vrp} instance with a 3-tuple $(N, K, W)$, where the element $w_{ij}$ of the matrix $W \in {\mathbb{R}^+}^{N \times N}$ codifies the cost for a vehicle to move from node $i$ to node $j$. A relevant example of cost matrix, which will be used for the rest of this article, is the distance matrix, which also exhibits the property $w_{ij}=w_{ji}\ \forall i,j$. Given these initial inputs and assuming without loss of generality that the depot node is the one with index $0$, the problem can be formally defined as: 
\begin{subequations}\label{eq:vrpdef}
\begin{align}
     \min_x & \sum_{i,j} w_{ij}x_{ij},\quad s.t.:\\
    & \sum_{j \in\delta^+(i)} x_{ij} = 1, \quad \forall i = 1, \dots, N-1,\label{eq:c1} \\
    & \sum_{j \in\delta^-(i)} x_{ji} = 1, \quad \forall i = 1, \dots, N-1,\label{eq:c2} \\
    & \sum_{j \in\delta^+(0)} x_{ij} = K, \label{eq:c3} \\
    & \sum_{j \in\delta^-(0)} x_{ji} = K, \label{eq:c4}\\
    & \sum_{i\in S}\sum_{j\notin S} x_{ij} \geq 1, \quad \forall S \in \left\{A \subset V \big | 2 \leq \left| A\right|\right\}, \label{eq:c5}
\end{align}
\end{subequations}
where $x=\{x_{ij}\}\in[0,1]^{N\times N}$ is a possible solution that has  the edge between a node $i$ and a node $j$ as part of that solution if $x_{ij}=1$ or without that edge if $x_{ij}=0$. The set of indices $\delta^+(i) = \left\{j \in V \big | (i, j) \in E\right\}$ is the set of outward connections from node $i$ and $\delta^-(j)= \left\{i \in V\ \big |\ (i, j) \in E\right\}$ the set of inward connections to $j$.
In this definition, equations \ref{eq:c1}--\ref{eq:c5} represent the constraints of the problem: \ref{eq:c1} and \ref{eq:c2} enforce that every non-depot node is visited exactly once guaranteeing that a vehicle reaching such a node must also leave it; \ref{eq:c3} and \ref{eq:c4} state that every vehicle must leave the depot and then return to it; lastly, equation \ref{eq:c5} prevents the occurrence of subtours, i.e., closed cycles that do not include the depot node, as shown in Figure~\ref{fig:subtour-example}.

Currently, the most straightforward approach to encode the constraints for the \gls{vrp} is to exclude the subtour-elimination constraint (Equation~\ref{eq:c5}), or, more generally, any inequality constraint that may require an excessive number of additional binary variables for the binarization of slack variables needed to transform the inequality into an equality. For instance, the subtour-elimination constraint requires a slack variable for every subset $S$ with at least $2$ nodes, resulting in an exponential number of additional variables as $N$ increases. This restricts its application to trivial problem instances. 

In order to solve \gls{vrp} by using quantum computers, we need to express it as a \gls{qubo} problem.
Our implementation of the constraints \ref{eq:c1}--\ref{eq:c4} follows~\citet{azad_solving_2023}, resulting in:
\begin{equation}
    \min_x x^TQ_\lambda x+g_\lambda^Tx
\end{equation}
over a vector variable $x$ that represents the flattened adjacency matrix of the solution assuming a fully connected graph without the self connections. In detail,
\begin{equation}
    x = \left[x_{0,1},x_{0,2},\dots,x_{0,N-1},x_{1,0},x_{1,2},\dots,x_{N-1,N-2}\right]\label{eq:bitstr}
\end{equation}
with $x_{i,j}=1$ if the connection $i\leadsto j$ is in the solution and $x_{i,j}=0$ otherwise. The terms $Q\lambda$ and $g_\lambda$ are defined as follows:
\begin{equation}
    Q_\lambda = \lambda\left(\sum_{k=1}^{N-1} z_{\delta^+(k)}z_{\delta^+(k)}^T+z_{\delta^-(k)}z_{\delta^-(k)}^T\right),
\end{equation}
\begin{equation}
    \begin{aligned}
    g_\lambda =w-2\lambda\Biggr[&\left(\sum_{k=1}^{N-1} z_{\delta^+(k)}^T+z_{\delta^-(k)}^T\right)\\
    &+K\left(z_{\delta^+(0)}^T+z_{\delta^-(0)}^T\right)\Biggr],
    \end{aligned}
\end{equation}
where the hyper-parameter $\lambda$ is the Lagrangian multiplier used to include the constraints into the objective function and \begin{equation}
z_{\delta^\pm(k)} = \left[z_{0,1},z_{0,2},\dots,z_{N-1,N-2}\right]
\end{equation}
are vectors with the same shape of $x$ that encode with $1$ the outward ($z_{\delta^+(k)}$) and inward ($z_{\delta^-(k)}$) connections of the $k$-th node, and with $0$ the other connections. 
By definition, $z_{i,j}=1 \iff i=k$ for $z_{\delta^+(k)}$ and $z_{i,j}=1 \iff j=k$ for $z_{\delta^-(k)}$. 

To solve \gls{vrp} in discrete circuit-based devices, we can express the \gls{qubo} problem with \gls{qaoa} as a cost Hamiltonian $H_C$ associated with the Ising model
\begin{equation}
    H_C = \sum_{i,j}J_{ij}Z_iZ_j + \sum_i h_i Z_i
\end{equation}
with the interaction coefficients depending on the \gls{qubo} terms in the following way:
    \begin{gather}
        J_{ij} = -\frac{{Q_\lambda}_{ij}}{4}, \\
        h_i = \frac{{g_\lambda}_i}{2} + \sum_j\frac{{Q_\lambda}_{ij}+{Q_\lambda}_{ji}}{4}.
    \end{gather}
By doing this, it is possible to discretize the dynamics with
\begin{equation}
\begin{aligned}
    U(\gamma, \beta) = \prod_{k=1}^P \Biggr(&\prod_i R_{X_i}(-\beta_k)\\
    &\prod_iR_{Z_i}(\gamma_kh_i) \prod_{ij}R_{Z_iZ_j}(\gamma_k J_{ij})\Biggr)
\end{aligned}
\end{equation}
where $\gamma=[\gamma_1,\dots,\gamma_k,\dots,\gamma_P]$ and $\beta=[\beta_1,\dots,\beta_k,\dots,\beta_P]$ are parameters and $P$ is a predefined hyperparameter.
By using this discrete evolution, the goal becomes minimizing the expectation value of the cost Hamiltonian:
\begin{equation}\label{eq:qaoa}
    \min_{\gamma, \beta} f(\gamma, \beta) =\bra{\psi(\gamma, \beta)}H_C\ket{\psi(\gamma, \beta)},
\end{equation}
with $\ket{\psi(\gamma,\beta)}=U(\gamma,\beta)\ket{+}$.

More details on \gls{qaoa} are reported in the appendix~\ref{sec:qaoa}.

\section{Results}
Our main contributions are the development of a novel initialization strategy for the \gls{qaoa} approach and the encoding of the subtour elimination constraint directly in the quantum optimization. The initialization strategy allows to increase the convergence probability of the quantum algorithm leading to higher probability of obtaining the correct solutions of the \gls{vrp}. Furthermore, we integrate the generic methodology of~\citet{montanez-barrera_unbalanced_2024} formulating a novel term for the \gls{qubo} problem specific for the \gls{vrp} successfully integrating the subtour constraint while avoiding the need for additional slack variables. In the following subsections, we discuss in detail the two contributions.

\subsection{Initialization strategy for QAOA}\label{sec:init}

Combinatorial optimization through \gls{qaoa} enables the application of quantum algorithms to real problems, even with \gls{nisq} devices, by leveraging its variational structure to learn a more noise-tolerant configuration of the system, capable of dealing with the problem at hand even in the presence of a reasonable amount of noise. However, this comes at the cost of dealing with the limitations of variational training protocols. In particular, \gls{qaoa} suffers from the \emph{barren plateau} phenomenon~\citep{anschuetz_unified_2025}, caused both by the large size of the system and by the use of a global expectation value required to estimate the total energy. A direct consequence is an increase in the number of training steps due to an exponentially vanishing gradient of the cost function, which can make training infeasible. Several solutions to this problem have been proposed: in~\citep{cerezo_cost_2021} it is shown how a cost function depending only on a local subset of qubits may exhibit better training properties, whereas in~\citep{yao_avoiding_2025} the authors provide a procedure for converting variational circuits from a unitary 2-design~\citep{mcclean_barren_2018} to a unitary 1-design using an additional ancilla register, improving gradient variance at the expense of more qubits; another well-known solution is to train the model layer-wise~\citep{skolik_layerwise_2021}, simplifying the cost landscape and improving conditioning, at the expense of a more expensive training procedure scaling with the number of layers.

\begin{figure*}[h!]
    \centering
    \includegraphics[width=0.7\linewidth]{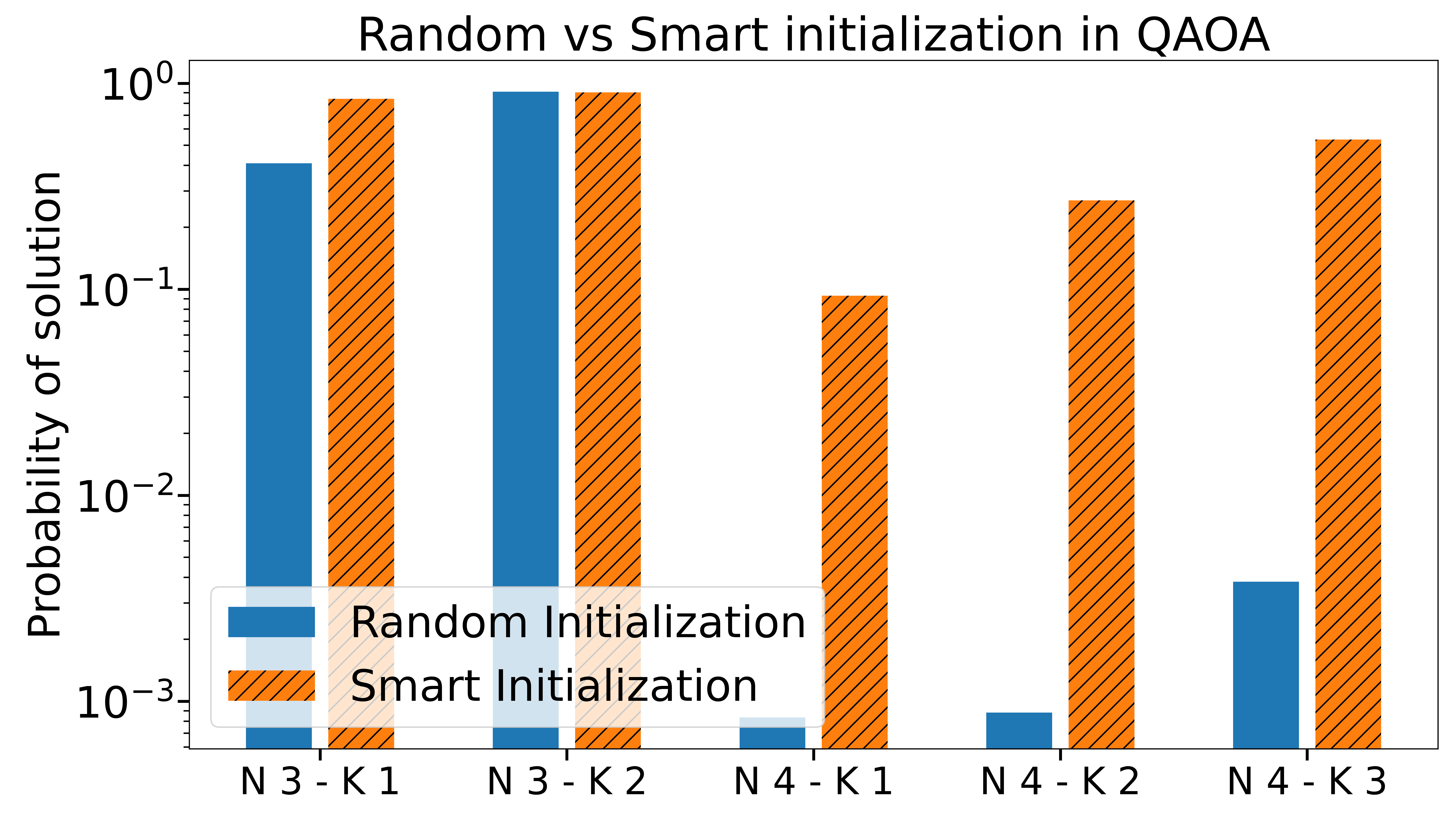}
    \caption{Comparison of the probability of sampling the optimal solution after executing \gls{qaoa} with a random initialization of the parameters (blue) and using our strategy (orange), evaluated for 50 different instances for each one of the 5 reported \gls{vrp} classes.}
    \label{fig:init-advantage}
\end{figure*}

An alternative class of plateau-mitigation techniques focuses directly on improving the optimization landscape conditioning through parameter initialization heuristics. Trotterized Quantum Annealing~\citep{sack2021quantum} uses physics-inspired adiabatic evolution schedules to systematically determine parameter values. It has also proven effective to transfer optimal parameter from smaller problem into the interested instance~\citep{zhou2020quantum, brandao2018foraging}. 
In contribution to these methods, we propose a novel initialization strategy based on a grid search over parameters, motivated by the bad conditioning of variational training under random initialization~\citep{mcclean_barren_2018}. In detail, we propose to evaluate the expected value of the energy of the system
\begin{equation}
\bra{\psi(\gamma, \beta)}H_C\ket{\psi(\gamma, \beta)}
\end{equation}
over a predetermined set of parameters that is choosen in a clever way. In particular, we evaluate all the configurations of $(\gamma, \beta)$ that satisfy $\gamma_k=\beta_k=\alpha_d\quad \forall k=1,\dots,P$ and for each value $\alpha_d=\frac{d\pi}{D}$ with $d=1\dots D$. The configuration with the lowest energy is used as the starting point for the \gls{qaoa} of Equation~\ref{eq:qaoa}.
This initialization requires $\mathcal{O}(D)$ additional circuit calls; $D$ is not directly proportional to the problem size and the procedure allows the training to start from a good approximation of the lowest-energy state.
Training from this starting point helps avoid barren plateaus and consequently increases the probability of sampling the solution at the end of training, as shown in Figure~\ref{fig:init-advantage}. In particular, we compared our approach with a random initialization of the parameters for five different classes of \gls{vrp} with three or four nodes and one to three vehicles. For each class, we trained the models over 50 different random instances and we report the average along the instances of the sum of the \gls{qaoa} output probabilities considering only the theoretical correct solutions for each instance. Notably, with higher value of $N$, the barren-plateau effects hinders the ability of updating the parameters, resulting in lower success probability using random initialization, while the smart initialization retains consistent performances even with an high qubit count.

\begin{figure*}[h!]
    \centering
    \includegraphics[width=0.9\linewidth]{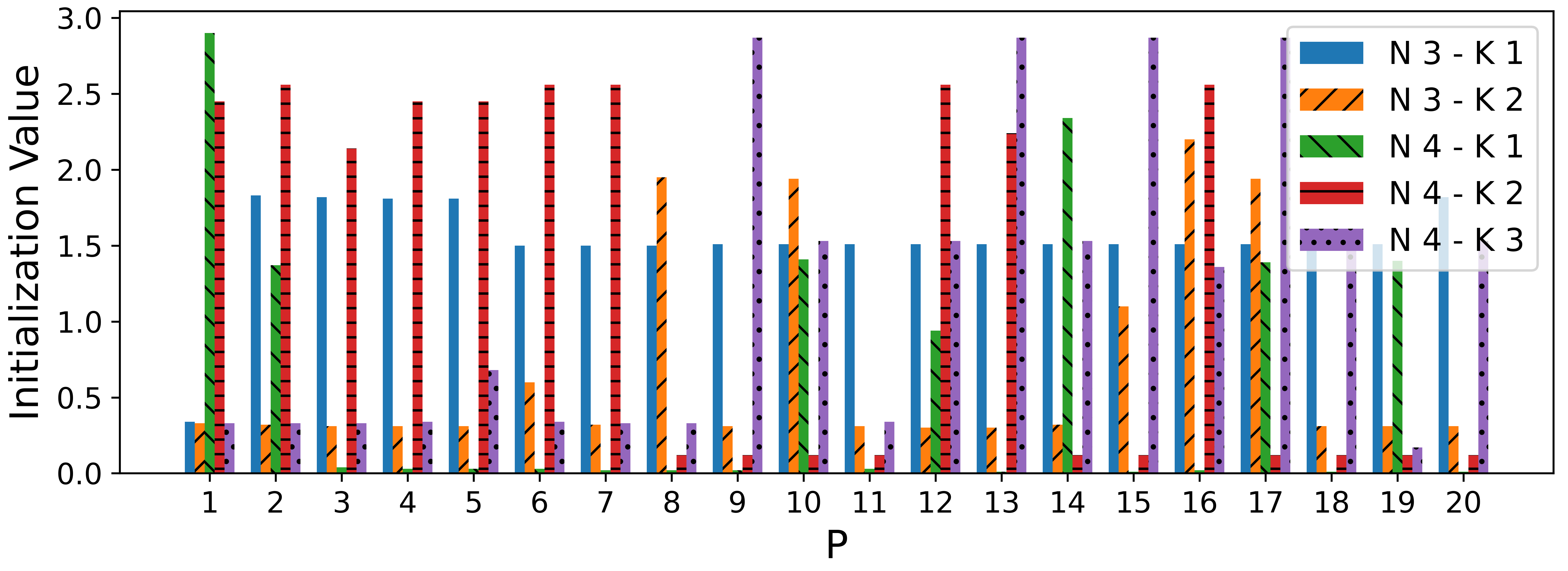}
    \caption{Starting point of the \gls{qaoa} training procedure (the optimal $\alpha_d$) obtained through our parameter initialization strategy for different circuit depths computed for $100$ different nodes configuration with coordinates component sampled from the distribution $\mathcal{N}(0,\sqrt{10})$. In all the cases the average starting value does not exhibit any kind of relationship with $P$, highlighting the necessity of spanning all the initialization space.}
    \label{fig:warm-up}
\end{figure*}

We executed tests to determine the effect of $P$ on the initial parameter search space, investigating a possible narrower interval $(0, B(P)]$, where $B(P)$ is a monotonic function of $P$. This follows the idea behind the \gls{qaoa} approximation of \gls{qa}, according to which more repetitions correspond to smaller time steps $\Delta_t$. However, as shown in Figure~\ref{fig:warm-up}, no relevant bias can be extrapolated as $P$ increases, ruling out the possibility of restricting the search for the optimal starting point to a smaller initialization interval $\mathcal{I}\subset (0, \pi]$.

\subsection{Encoding the subtour elimination constraint}\label{sec:subtour}
The exclusion of the subtour elimination constraint (Equation~\ref{eq:c5}), being hard to implement due to the additional variables needed to express them in the \gls{qubo} formalism, requires a round of classical post-processing aimed at filtering the solutions presenting subtours at the end of the algorithm~\citep{azad_solving_2023}. This process, despite being less demanding from a computational standpoint, has still an exponential scaling in its complexity. Therefore, as the problem size increases, the eventual advantage provided by using \gls{qaoa} is limited.

To achieve a true end-to-end quantum solution to the problem, all the constraints must be included in the system Hamiltonian, an approach that is usually only feasible with \gls{qa} solvers~\citep{kim_quantum_2025}, as they provide larger quantum registers. 

\begin{figure*}[h!]
    \centering
    \includegraphics[width=0.7\linewidth]{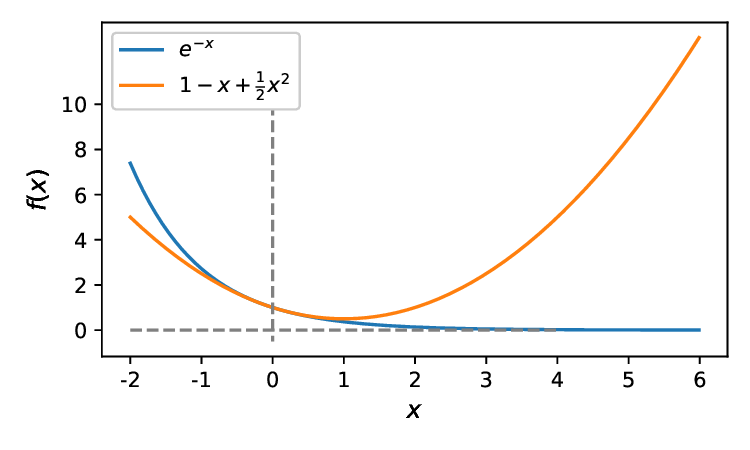}
    \caption{Since it is not known a priori how strongly the inequality constraint should weigh on the cost function (depending on the unknown position of the optimum), a narrow quadratic form may penalize it excessively, while a wider quadratic form may be unable to enforce the constraint.}
    \label{fig:unbalanced-approximation}
\end{figure*}

Building on the work of Montañez-Barrera et al. 2024~\citep{montanez-barrera_unbalanced_2024}, where the authors propose to encode inequality constraints with the second order Taylor expansion of the constraint of the exponential function, resulting in the following relaxed optimization problem: 
\begin{equation}
\begin{aligned}
\begin{aligned}
    \min_x &\ f(x) \\
    & h(x)\geq 0 & 
\end{aligned} &\longrightarrow \min_x f(x) + \lambda e^{-h(x)}\\
&\longrightarrow \min_x f(x) - \lambda_1h(x)+\lambda_2\frac{1}{2}h(x)^2.
\end{aligned}
\end{equation} 
Doing so allows the encoding of inequalities without requiring additional variables, at the expense of introducing two additional hyper-parameters $\lambda_{1}$ and $\lambda_{2}$, used to shape the approximating quadratic form, as shown in Figure~\ref{fig:unbalanced-approximation}.
For every subset $S \in \left\{A \subset V\ \big |\ 2 \leq \left| A\right|\right\}$ we define the vector $\gamma_S$, using the same indices convention as the binary variable $x$, as the set-transition vector: \begin{equation}
    \gamma_S^{(ij)} = \mathbbm{1} _{\left\{ i\in S,\ j\notin S\right\}}, 
\end{equation} 
with $\mathbbm{1}_{\left\{\cdot\right\}}$ denoting the indicator function. Using this notation, our subtour elimination constraint can be expressed as \begin{equation}
    h_S(x) = \gamma_S^Tx - 1 \geq 0.
\end{equation}
In the application of the Taylor expansion of $e^{-\gamma_S^Tx+1}$, by considering $\gamma_S^Tx$ as the variable, is it possible to derive the approximation \begin{equation}
    e^{-\gamma_S^Tx+1} \approx e-e\cdot \gamma_S^Tx + \frac{e}{2}x^T\gamma_S\gamma_S^Tx
\end{equation}
by considering the contribution of every subset $S$ and the introduction of the Lagrangian penalties, we retrieve the final \gls{qubo} problem constraint \begin{equation}\label{eq:qvrp}
\begin{aligned}
    \min_x\ &x^TQ_\lambda x+g_\lambda^Tx +  \lambda_{\text{0}}\underbrace{e|S|}_{c_\text{subtour}}\\
    &-\lambda_{\text{1}}\underbrace{e\sum_{S\in\mathcal{S}}\gamma_S^T}_{g_\text{subtour} }x +\lambda_{\text{2}}x^T\underbrace{\left(\frac{e}{2}\sum_{S\in\mathcal{S}}\gamma_S\gamma_S^T\right)}_{Q_\text{subtour}}x 
\end{aligned}
\end{equation}
where $\left(\lambda, \lambda_1, \lambda_2\right)$ represents the hyper-parameter vector that shapes the \gls{qubo} encoding of the problem. The constant term, being irrelevant to the optimization procedure, can be omitted.

The final system then is dependent on a total of three hyper-parameters: the weight of the first four constraints (Equations~\ref{eq:c1} to~\ref{eq:c4}) $\lambda$, which can be either a scalar value or a function of the cost matrix; the weight of the linear and quadratic term of the subtour constraint $\lambda_1$ and $\lambda_2$. The ideal value of the hyper-parameters is obtained using Optuna~\citep{akiba_optuna_2019}, with the hyper-parameter space defined as in the last row of Table~\ref{tab:best-configs}. 

\begin{figure*}[h!]
    \centering
    \includegraphics[width=\linewidth]{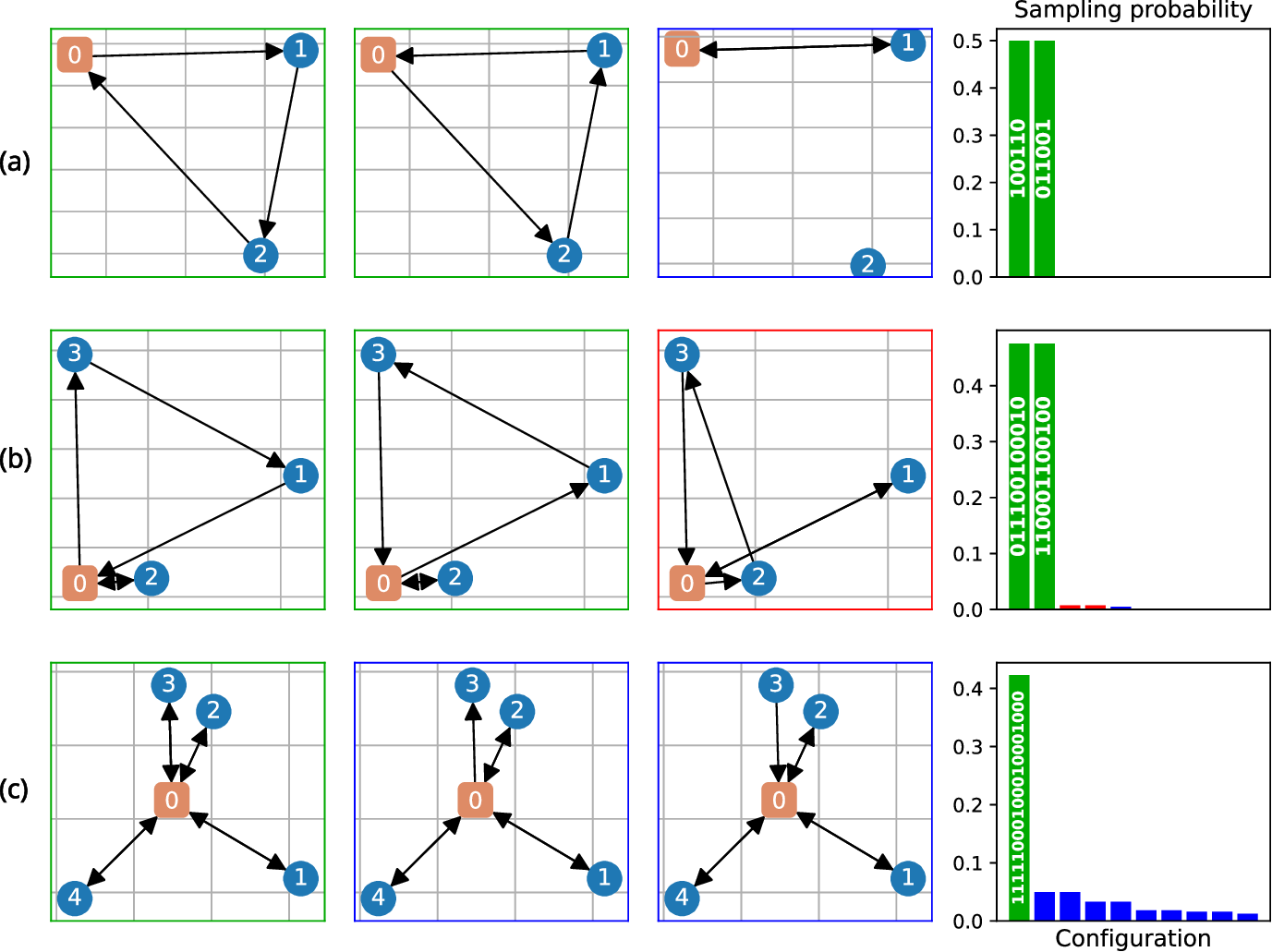}
    
    \caption{Solution generated for the case $VRP(3,1)$ (a), $VRP(4,2)$ (b) and $VRP(5, 4)$ (c). The three most probable configuration are reported on the left, and the distribution of the top 10 most probable states are reported on the right, following the notation provided in Equation~\ref{eq:bitstr}. The optimal solutions are highlighted in green, the non-optimal feasible solutions in red and the unfeasible solutions in blue. \gls{qaoa} is capable of isolating the optimal solution (which correspond to the first two configuration in (a) and (b), and the only admissible solution in (c)) from the rest of the configurations. }
    \label{fig:results}
\end{figure*}

Using the optimal hyper-parameter configuration, reported in Table~\ref{tab:best-configs}, we achieve a correct end-to-end prediction of the \gls{vrp} on all the configurations considered. In particular, we report the sum of the \gls{qaoa} output probabilities for the theoretically correct solutions. In all cases, the \gls{qaoa} output probabilities are concentrated on the correct ones, in the sense that the highest predicted configurations were the theoretically correct solutions of the \gls{vrp}. It should be noted that with $N=3$, the subtour constraint is automatically satisfied, since the condition in Equations~\ref{eq:c1}-\ref{eq:c4} are met. Therefore, as we can observe in Figure~\ref{fig:results}a, despite the introduction of unnecessary terms, the generality of this encoding is proved for every case.
\begin{table*}
\centering
\caption{Optimal hyper-parameters configuration found, alongside the probability of sampling the solution configuration. In the last row we report also the search space associated to Equation~\ref{eq:qvrp}. The search space for $\lambda$, $\lambda_1$ and $\lambda_2$ is the same. Since the original \gls{qubo} cost function components $(Q_\lambda, g_\lambda)$ depends on the weight matrix, the first penalty coefficient $\lambda$ can be scaled with the 1-norm (sum) or the infinity-norm (max) of the weight matrix $W$. The optimal combination of hyper-parameter is obtained through a grid-search using Optuna.}\label{tab:best-configs}
\renewcommand{\arraystretch}{1.3}
\begin{tabular}{cccccc}
\toprule
\textbf{Configuration} & \multirow{2}{*}{$\lambda$ scaling} & \multirow{2}{*}{$\lambda$} & \multirow{2}{*}{$\lambda_1$} & \multirow{2}{*}{$\lambda_2$} & Solution \\
$(N,K,P)$ & & & & & probability \\ \midrule
$(3,1,6)$ & \texttt{max}&\texttt{4.5}&\texttt{1}&\texttt{1} & $~\approx
 0.9999$\\
$(3,2,6)$ & \texttt{max}&\texttt{3}&\texttt{1}&\texttt{1} & $~\approx
 0.9999$\\
$(3,1,20)$ & \texttt{max}&\texttt{1}&\texttt{1.5}&\texttt{3} & $\approx 0.4969$\\
$(4,2,20)$ & \texttt{max}&\texttt{6}&\texttt{1}&\texttt{2} & $\approx 0.9506$\\
$(4,3,20)$ & \texttt{max}&\texttt{3.5}&\texttt{1.5}&\texttt{1.5} & $\approx 0.9975 $\\
$(5,4,24)$ & \texttt{max}&\texttt{7.5}&\texttt{1}&\texttt{1} &$\approx 0.423$\\
\bottomrule
\textbf{Search space:} & \texttt{[flat, max, sum]} & \multicolumn{3}{c}{\texttt{[1, 1.5, 2, \dots, 9.5]}} & \\
\bottomrule
\end{tabular}
\end{table*}
With this encoding, we also manage to solve the problem for the cases $N=4$ and $N=5$. From the histograms reported in Figures~\ref{fig:results}b--\ref{fig:results}c, it is possible to appreciate how, given the right encoding, \gls{qaoa} can produce a sharp probability distribution centered on the optimal solution.

Further experiments on the generality of the hyper-parameters show that, within the same class of \gls{vrp} instances, the same hyper-parameters (or a nearby point in the configuration space) might be applicable to different instances. However, changing the class of the problem (modifying either $N$ or $K$) requires a new hyper-parameter configuration.

\section{Conclusions}\label{sec:conclusion}
\gls{vrp} is a sensible topic with tangible repercussions on the economy and society. Therefore, improvements on this field are desirable and can lead to mitigation of costs and environmental preservation. In parallel, quantum computing and quantum optimization represent the frontiers in computation, allowing to tackle some classes of NP-hard problems in a tractable way as the \gls{vrp} without the soubtour constraints. However, ignoring the subtour constraints means that the produced solutions needs to be filtered in post-processing to avoid unfeasible solutions. This post-processing is a computational burden that needs to be considered for practical implementations of classical or quantum optimization implementations to solve \gls{vrp}. The usual approaches to face \gls{vrp} using quantum techniques involve \gls{qa} on quantum annealers or \gls{qaoa} on circuit-based devices.

In this work, we improved the \gls{qaoa} approach to solve \gls{vrp} by introducing two novel techniques in this field. In particular, we successfully implemented a clever initialization of the \gls{qaoa} parameters and we integrated the subtour constraints within the quantum computation without requiring additional qubits and classical post-processing. The proposed initialization strategy does not make any assumptions about the problem instance. It is therefore an effective method to improve the success rate of the training procedure and to mitigate barren plateaus, one of the biggest drawbacks in quantum optimization. Furthermore, compared with the standard slack-variable formulation, the proposed encoding reduces the number of required qubits by $\mathcal{O}\left(2^N\right)$ for instances with $N$, while achieving a high probability of sampling.

The presented methods are general in scope. With a different derivation of the inequalities, the unbalanced exponential codification of the constraints can also be applied to other problems, as shown in~\citep{montanez-barrera_unbalanced_2024}. This enables a more qubit-efficient solution of optimization problems on quantum computers, at the expense of additional hyper-parameters that have to be tuned. 

At best of our knowledge, we were the first to achieve an effective initialization of the \gls{qaoa} parameters within the \gls{vrp} context that is independent on the problem size. Moreover, we were the first to adopt the encoding of the subtour inequalities of \gls{vrp} directly into the \gls{qaoa} optimization.

We validated our contributes on different classes of \gls{vrp} and showed how they can be used effectively in this context. In particular, we reported that our proposed parameters initialization is statistically better than a random initialization of the parameters and that our integration of the subtour constraint allows to find the correct optimal solutions of \gls{vrp} for different classes of problems with $N$ from 3 to 5.

Some limitations of the present study should be acknowledged. Due to computational demand, the experiments were restricted to small instances and carried out on noiseless simulators, so the effect of noise and decoherence on the proposed encoding has not been fully characterized. 
Moreover, conversion into \gls{qubo} formalism of the constraints is highly sensible to the choice of the hyper-parameters, which determines whether the optimal feasible solution corresponds to the ground state of the resulting Hamiltonian. 

Given the highly structured nature of combinatorial problems such as the \gls{vrp}, it is plausible that these hyper-parameters can be determined through a metaheuristic based on problem invariants. Future works may therefore focus on finding hyper-parameter configurations that are either general or dependent on the instance class (i.e., as a function of $N$ and $K$). Such configurations could serve as a starting point for the hyper-parameter search, which represents the bottleneck of any \gls{qaoa} approach requiring a conversion of the original problem to \gls{qubo}. 

Despite being limited in problem size by current \gls{nisq} restrictions, which also prevent seamless integration of inequality constraints, this work shows that the required qubit-register size can be reduced by engineering such constraints, improving the efficiency of the \gls{qubo} conversion. This paves the way for future implementations of \gls{vrp} on fault-tolerant quantum computers, which may become a viable alternative or complement to classical routing methods. More broadly, we expect careful constraint engineering to remain a key ingredient in making quantum optimization competitive for industrially relevant logistics problems.

\backmatter

\bmhead{Competing Interests}
The authors declare no competing interests.

\bmhead{Acknowledgments}
S.M. and F.C. acknowledge financial support from the PNRR MUR project PE0000023-NQSTI and from Fondazione Cassa di Risparmio di Firenze.

\bmhead{Data Availability}
The data used in this study is dynamically generated using the modality reported in the article. For further clarification, feel free to contact the corresponding author. 
\bmhead{Code Availability}
The underlying code for this study is not publicly available, but may be made available to qualified researchers on reasonable request from the corresponding author. 
\bmhead{Author Contributions}
E.C. carried out the algorithm implementation and numerical experiments based on the initial work provided by M.D.S. and with the theoretical support of S.M.; E.C., S.M. and F.C. conceived the methodology, while F.C. proposed and supervised the project. All authors contributed to the discussion, analysis of the results, and the writing of the manuscript.

\bibliography{biblioArxiv}

@article{montanez-barrera_unbalanced_2024,
    title = {Unbalanced penalization: a new approach to encode inequality constraints of combinatorial problems for quantum optimization algorithms},
    volume = {9},
    issn = {2058-9565},
    shorttitle = {Unbalanced penalization},
    url = {https://doi.org/10.1088/2058-9565/ad35e4},
    doi = {10.1088/2058-9565/ad35e4},
    language = {en},
    number = {2},
    journal = {Quantum Science and Technology},
    publisher = {IOP Publishing},
    author = {Montañez-Barrera, J A and Willsch, Dennis and Maldonado-Romo, A and Michielsen, Kristel},
    month = apr,
    year = {2024},
    pages = {025022},
}

@misc{irie_quantum_2019,
    title = {Quantum {Annealing} of {Vehicle} {Routing} {Problem} with {Time}, {State} and {Capacity}},
    url = {http://arxiv.org/abs/1903.06322},
    doi = {10.48550/arXiv.1903.06322},
    
    publisher = {arXiv},
    author = {Irie, Hirotaka and Wongpaisarnsin, Goragot and Terabe, Masayoshi and Miki, Akira and Taguchi, Shinichirou},
    month = mar,
    year = {2019},
    note = {arXiv:1903.06322 [quant-ph]},
}

@article{borowski_new_2020,
    title = {New {Hybrid} {Quantum} {Annealing} {Algorithms} for {Solving} {Vehicle} {Routing} {Problem}},
    volume = {12142},
    url = {https://pmc.ncbi.nlm.nih.gov/articles/PMC7304725/},
    doi = {10.1007/978-3-030-50433-5\_42},
    
    journal = {Computational Science – ICCS 2020},
    author = {Borowski, Michał and Gora, Paweł and Karnas, Katarzyna and Błajda, Mateusz and Król, Krystian and Matyjasek, Artur and Burczyk, Damian and Szewczyk, Miron and Kutwin, Michał},
    month = may,
    year = {2020},
    pages = {546--561},
}

@misc{huang_solving_2025,
    title = {Solving {Capacitated} {Vehicle} {Routing} {Problem} with {Quantum} {Alternating} {Operator} {Ansatz} and {Column} {Generation}},
    url = {http://arxiv.org/abs/2503.17051},
    doi = {10.48550/arXiv.2503.17051},
    
    publisher = {arXiv},
    author = {Huang, Wei-hao and Matsuyama, Hiromichi and Yamashiro, Yu},
    month = mar,
    year = {2025},
    note = {arXiv:2503.17051 [quant-ph]},
}

@misc{azfar_quantum-assisted_2026,
    title = {Quantum-{Assisted} {Vehicle} {Routing}: {Realizing} {QAOA}-based {Approach} on {Gate}-{Based} {Quantum} {Computer}},
    shorttitle = {Quantum-{Assisted} {Vehicle} {Routing}},
    url = {http://arxiv.org/abs/2505.01614},
    doi = {10.48550/arXiv.2505.01614},
    
    publisher = {arXiv},
    author = {Azfar, Talha and Raisuddin, Osama Muhammad and Ke, Ruimin and Holguin-Veras, Jose},
    month = jan,
    year = {2026},
    note = {arXiv:2505.01614 [quant-ph]},
}

@article{wang_xy_2020,
    title = {\${XY}\$ mixers: {Analytical} and numerical results for the quantum alternating operator ansatz},
    volume = {101},
    shorttitle = {\${XY}\$ mixers},
    url = {https://link.aps.org/doi/10.1103/PhysRevA.101.012320},
    doi = {10.1103/PhysRevA.101.012320},
    number = {1},
    
    journal = {Physical Review A},
    publisher = {American Physical Society},
    author = {Wang, Zhihui and Rubin, Nicholas C. and Dominy, Jason M. and Rieffel, Eleanor G.},
    month = jan,
    year = {2020},
    pages = {012320},
}

@article{azad_solving_2023,
    title = {Solving {Vehicle} {Routing} {Problem} {Using} {Quantum} {Approximate} {Optimization} {Algorithm}},
    volume = {24},
    issn = {1558-0016},
    url = {https://ieeexplore.ieee.org/abstract/document/9774961},
    doi = {10.1109/TITS.2022.3172241},
    number = {7},
    
    journal = {IEEE Transactions on Intelligent Transportation Systems},
    author = {Azad, Utkarsh and Behera, Bikash K. and Ahmed, Emad A. and Panigrahi, Prasanta K. and Farouk, Ahmed},
    month = jul,
    year = {2023},
    pages = {7564--7573},
}

@article{braess_paradox_2005,
    title = {On a {Paradox} of {Traffic} {Planning}},
    volume = {39},
    issn = {0041-1655, 1526-5447},
    url = {https://pubsonline.informs.org/doi/10.1287/trsc.1050.0127},
    doi = {10.1287/trsc.1050.0127},
    language = {en},
    number = {4},
    
    journal = {Transportation Science},
    author = {Braess, Dietrich and Nagurney, Anna and Wakolbinger, Tina},
    month = nov,
    year = {2005},
    pages = {446--450},
}

@article{cerezo_cost_2021,
    title = {Cost {Function} {Dependent} {Barren} {Plateaus} in {Shallow} {Parametrized} {Quantum} {Circuits}},
    volume = {12},
    issn = {2041-1723},
    url = {http://arxiv.org/abs/2001.00550},
    doi = {10.1038/s41467-021-21728-w},
    number = {1},
    
    journal = {Nature Communications},
    author = {Cerezo, M. and Sone, Akira and Volkoff, Tyler and Cincio, Lukasz and Coles, Patrick J.},
    month = mar,
    year = {2021},
    note = {arXiv:2001.00550 [quant-ph]},
    pages = {1791},
}

@article{yao_avoiding_2025,
    title = {Avoiding {Barren} {Plateaus} with {Entanglement}},
    volume = {111},
    issn = {2469-9926, 2469-9934},
    url = {http://arxiv.org/abs/2406.03748},
    doi = {10.1103/PhysRevA.111.022426},
    number = {2},
    
    journal = {Physical Review A},
    author = {Yao, Yuhan and Hasegawa, Yoshihiko},
    month = feb,
    year = {2025},
    note = {arXiv:2406.03748 [quant-ph]},
    pages = {022426},
}

@article{skolik_layerwise_2021,
    title = {Layerwise learning for quantum neural networks},
    volume = {3},
    issn = {2524-4906, 2524-4914},
    url = {http://arxiv.org/abs/2006.14904},
    doi = {10.1007/s42484-020-00036-4},
    number = {1},
    
    journal = {Quantum Machine Intelligence},
    author = {Skolik, Andrea and McClean, Jarrod R. and Mohseni, Masoud and Smagt, Patrick van der and Leib, Martin},
    month = jun,
    year = {2021},
    note = {arXiv:2006.14904 [quant-ph]},
    pages = {5},
}

@article{feld_hybrid_2019,
    title = {A {Hybrid} {Solution} {Method} for the {Capacitated} {Vehicle} {Routing} {Problem} {Using} a {Quantum} {Annealer}},
    volume = {6},
    issn = {2297-198X},
    url = {https://www.frontiersin.org/journals/ict/articles/10.3389/fict.2019.00013/full},
    doi = {10.3389/fict.2019.00013},
    language = {English},
    
    journal = {Frontiers in ICT},
    publisher = {Frontiers},
    author = {Feld, Sebastian and Roch, Christoph and Gabor, Thomas and Seidel, Christian and Neukart, Florian and Galter, Isabella and Mauerer, Wolfgang and Linnhoff-Popien, Claudia},
    month = jun,
    year = {2019},
}

@article{xie_feasibility-preserved_2024,
    title = {A feasibility-preserved quantum approximate solver for the {Capacitated} {Vehicle} {Routing} {Problem}},
    volume = {23},
    issn = {1573-1332},
    url = {https://doi.org/10.1007/s11128-024-04497-5},
    doi = {10.1007/s11128-024-04497-5},
    language = {en},
    number = {8},
    
    journal = {Quantum Information Processing},
    author = {Xie, Ningyi and Lee, Xinwei and Cai, Dongsheng and Saito, Yoshiyuki and Asai, Nobuyoshi and Lau, Hoong Chuin},
    month = jul,
    year = {2024},
    pages = {291},
}

@misc{farhi_quantum_2014,
    title = {A {Quantum} {Approximate} {Optimization} {Algorithm}},
    url = {http://arxiv.org/abs/1411.4028},
    doi = {10.48550/arXiv.1411.4028},
    
    publisher = {arXiv},
    author = {Farhi, Edward and Goldstone, Jeffrey and Gutmann, Sam},
    month = nov,
    year = {2014},
    note = {arXiv:1411.4028 [quant-ph]},
}

@article{morita_mathematical_2008,
    title = {Mathematical {Foundation} of {Quantum} {Annealing}},
    volume = {49},
    issn = {0022-2488, 1089-7658},
    url = {http://arxiv.org/abs/0806.1859},
    doi = {10.1063/1.2995837},
    number = {12},
    
    journal = {Journal of Mathematical Physics},
    author = {Morita, Satoshi and Nishimori, Hidetoshi},
    month = dec,
    year = {2008},
    note = {arXiv:0806.1859 [quant-ph]},
    pages = {125210},
}

@article{laporte_branch_1983,
    title = {A branch and bound algorithm for the capacitated vehicle routing problem},
    volume = {5},
    issn = {1436-6304},
    url = {https://doi.org/10.1007/BF01720015},
    doi = {10.1007/BF01720015},
    language = {en},
    number = {2},
    
    journal = {Operations-Research-Spektrum},
    author = {Laporte, G. and Nobert, Y.},
    month = jun,
    year = {1983},
    pages = {77--85},
}

@article{carpaneto_branch_1989,
    title = {A branch and bound algorithm for the multiple depot vehicle scheduling problem},
    volume = {19},
    copyright = {Copyright © 1989 Wiley Periodicals, Inc., A Wiley Company},
    issn = {1097-0037},
    url = {https://onlinelibrary.wiley.com/doi/abs/10.1002/net.3230190505},
    doi = {10.1002/net.3230190505},
    abstract = {The Vehicle Scheduling Problem concerns the assigning of a set of time-tabled trips to vehicles so as to minimize a given cost function. We consider the NP-hard Multiple Depot case in which, in addition, one has to assign vehicles to depots. Different lower bounds based on assigment relaxation and on connectivity constraints are presented and combined in an effective bounding procedure. A strong dominance procedure derived from new dominance criteria also described. A branch and bound algorithm is finally proposed. Computational results are given.},
    language = {en},
    number = {5},
    
    journal = {Networks},
    author = {Carpaneto, G. and Dell'amico, M. and Fischetti, M. and Toth, P.},
    year = {1989},
    _note = {\_eprint: https://onlinelibrary.wiley.com/doi/pdf/10.1002/net.3230190505},
    pages = {531--548},
}

@misc{liu_heuristics_2023,
    title = {Heuristics for {Vehicle} {Routing} {Problem}: {A} {Survey} and {Recent} {Advances}},
    shorttitle = {Heuristics for {Vehicle} {Routing} {Problem}},
    url = {http://arxiv.org/abs/2303.04147},
    doi = {10.48550/arXiv.2303.04147},
    language = {en},
    
    publisher = {arXiv},
    author = {Liu, Fei and Lu, Chengyu and Gui, Lin and Zhang, Qingfu and Tong, Xialiang and Yuan, Mingxuan},
    month = mar,
    year = {2023},
    note = {arXiv:2303.04147 [cs.AI]},
}

@article{wierichs_general_2022,
	title = {General parameter-shift rules for quantum gradients},
	volume = {6},
	url = {https://quantum-journal.org/papers/q-2022-03-30-677/},
	doi = {10.22331/q-2022-03-30-677},
	language = {en-GB},
	
	journal = {Quantum},
	publisher = {Verein zur Förderung des Open Access Publizierens in den Quantenwissenschaften},
	author = {Wierichs, David and Izaac, Josh and Wang, Cody and Lin, Cedric Yen-Yu},
	month = mar,
	year = {2022},
	pages = {677},
}

@book{arora2006computational,
  author    = {Arora, Sanjeev and Barak, Boaz},
  title     = {Computational Complexity: A Modern Approach},
  publisher = {Cambridge University Press},
  address   = {Cambridge},
  year      = {2009},
  isbn      = {978-0-521-42426-4},
  url       = {https://theory.cs.princeton.edu/complexity/book.pdf}
}

@misc{anschuetz_unified_2025,
    title = {A {Unified} {Theory} of {Quantum} {Neural} {Network} {Loss} {Landscapes}},
    url = {http://arxiv.org/abs/2408.11901},
    doi = {10.48550/arXiv.2408.11901},
    
    publisher = {arXiv},
    author = {Anschuetz, Eric R.},
    month = feb,
    year = {2025},
    note = {arXiv:2408.11901 [quant-ph]},
}

@inproceedings{akiba_optuna_2019,
    address = {New York, NY, USA},
    series = {{KDD} '19},
    title = {Optuna: {A} {Next}-generation {Hyperparameter} {Optimization} {Framework}},
    isbn = {978-1-4503-6201-6},
    shorttitle = {Optuna},
    url = {https://dl.acm.org/doi/10.1145/3292500.3330701},
    doi = {10.1145/3292500.3330701},
    
    booktitle = {Proceedings of the 25th {ACM} {SIGKDD} {International} {Conference} on {Knowledge} {Discovery} \& {Data} {Mining}},
    publisher = {Association for Computing Machinery},
    author = {Akiba, Takuya and Sano, Shotaro and Yanase, Toshihiko and Ohta, Takeru and Koyama, Masanori},
    month = jul,
    year = {2019},
    pages = {2623--2631},
}

@article{kumar_survey_2012,
	title = {A {Survey} on the {Vehicle} {Routing} {Problem} and {Its} {Variants}},
	volume = {4},
	copyright = {https://creativecommons.org/licenses/by/4.0/},
	url = {https://www.scirp.net/journal/paperinformation?paperid=19355},
	doi = {10.4236/iim.2012.43010},
	language = {en},
	number = {3},
	
	journal = {Intelligent Information Management},
	publisher = {Scientific Research Publishing},
	author = {Kumar, Suresh Nanda and Panneerselvam, Ramasamy},
	month = may,
	year = {2012},
	pages = {66--74},
}

@article{abdalkareem_healthcare_2021,
	title = {Healthcare scheduling in optimization context: a review},
	volume = {11},
	issn = {2190-7188},
	shorttitle = {Healthcare scheduling in optimization context},
	doi = {10.1007/s12553-021-00547-5},
	language = {eng},
	number = {3},
	journal = {Health and Technology},
	author = {Abdalkareem, Zahraa A. and Amir, Amiza and Al-Betar, Mohammed Azmi and Ekhan, Phaklen and Hammouri, Abdelaziz I.},
	year = {2021},
	pages = {445--469},
}

@article{crama_combinatorial_1997,
	title = {Combinatorial optimization models for production scheduling in automated manufacturing systems},
	volume = {99},
	issn = {0377-2217},
	url = {https://www.sciencedirect.com/science/article/pii/S0377221796003888},
	doi = {10.1016/S0377-2217(96)00388-8},
	number = {1},
	
	journal = {European Journal of Operational Research},
	author = {Crama, Yves},
	month = may,
	year = {1997},
	pages = {136--153},
}

@article{boria_survey_2011,
	title = {A survey on combinatorial optimization in dynamic environments},
	volume = {45},
	copyright = {© EDP Sciences, ROADEF, SMAI 2011},
	issn = {0399-0559, 1290-3868},
	url = {https://www.rairo-ro.org/articles/ro/abs/2011/03/ro110015/ro110015.html},
	doi = {10.1051/ro/2011114},
	language = {en},
	number = {3},
	
	journal = {RAIRO - Operations Research},
	publisher = {EDP Sciences},
	author = {Boria, Nicolas and Paschos, Vangelis T.},
	month = jul,
	year = {2011},
	pages = {241--294},
}

@article{kim_quantum_2025,
	title = {Quantum annealing for combinatorial optimization: a benchmarking study},
	volume = {11},
	copyright = {2025 The Author(s)},
	issn = {2056-6387},
	shorttitle = {Quantum annealing for combinatorial optimization},
	url = {https://www.nature.com/articles/s41534-025-01020-1},
	doi = {10.1038/s41534-025-01020-1},
	language = {en},
	number = {1},
	
	journal = {npj Quantum Information},
	publisher = {Nature Publishing Group},
	author = {Kim, Seongmin and Ahn, Sang-Woo and Suh, In-Saeng and Dowling, Alexander W. and Lee, Eungkyu and Luo, Tengfei},
	month = may,
	year = {2025},
	pages = {77},
}

@article{laporte_traveling_1992,
	title = {The traveling salesman problem: {An} overview of exact and approximate algorithms},
	volume = {59},
	issn = {0377-2217},
	shorttitle = {The traveling salesman problem},
	url = {https://www.sciencedirect.com/science/article/pii/037722179290138Y},
	doi = {10.1016/0377-2217(92)90138-Y},
	number = {2},
	
	journal = {European Journal of Operational Research},
	author = {Laporte, Gilbert},
	month = jun,
	year = {1992},
	pages = {231--247},
}

@incollection{nemhauser_general_1988,
	title = {General {Algorithms}},
	isbn = {978-1-118-62737-2},
	_url = {https://onlinelibrary.wiley.com/doi/abs/10.1002/9781118627372.ch11},
	doi = {10.1002/9781118627372.ch11},
	abstract = {This chapter discusses approaches for finding an optimal, or e-approximate, solution of the linear integer programming problem. It gives a fractional cutting-plane algorithm (FCPA) for general integer programs that uses C-G inequalities. The chapter presents a fractional cutting-plane algorithm (FCPA) for pure-integer programs that uses Gomory cuts. The main result is that the Gomory FCPA is finitely convergent. The chapter then extends the algorithm to mixed-integer programs and describes a primal cutting-plane algorithm for pure-integer programs. It progresses by generating adjacent extreme points of the convex hull of feasible integral solutions. Enumerative methods go under the general names of branch-and-bound, implicit enumeration, and divide-and conquer. The latter term is frequently used in the computer science literature, and the first two terms are commonly used in the mathematical programming/operations research literature.},
	language = {en},
	
	booktitle = {Integer and {Combinatorial} {Optimization}},
	publisher = {John Wiley \& Sons, Ltd},
	author = {Nemhauser, George and Wolsey, Laurence},
	year = {1988},
	_note = {Section: II.4
\_eprint: https://onlinelibrary.wiley.com/doi/pdf/10.1002/9781118627372.ch11},
	keywords = {branch-and-bound, divide-and conquer, fractional cutting-plane algorithm (FCPA), general algorithms, Gomory cuts, implicit enumeration, linear integer programming problem, mixed-integer programs},
	pages = {349--382},
    address = {Hoboken, NJ}
}

@incollection{korte_knapsack_2008,
	address = {Berlin, Heidelberg},
    author = {Bernhard, Korte and Jens, Vygen},
	title = {The {Knapsack} {Problem}},
	isbn = {978-3-540-71844-4},
	url = {https://doi.org/10.1007/978-3-540-71844-4\_17},
	doi = {10.1007/978-3-540-71844-4\_17},
	language = {en},
	
	booktitle = {Combinatorial {Optimization}: {Theory} and {Algorithms}},
	publisher = {Springer},
	editor = {Korte, Bernhard and Vygen, Jens},
	year = {2008},
	pages = {439--448},
}

@misc{kingma_adam_2017,
    title = {Adam: {A} {Method} for {Stochastic} {Optimization}},
    shorttitle = {Adam},
    url = {http://arxiv.org/abs/1412.6980},
    doi = {10.48550/arXiv.1412.6980},
    
    publisher = {arXiv},
    author = {Kingma, Diederik P. and Ba, Jimmy},
    month = jan,
    year = {2017},
    note = {arXiv:1412.6980 [cs]},
}

@misc{bergholm_pennylane_2018,
    title = {{PennyLane}: {Automatic} differentiation of hybrid quantum-classical computations},
    shorttitle = {{PennyLane}},
    url = {https://ui.adsabs.harvard.edu/abs/2018arXiv181104968B},
    doi = {10.48550/arXiv.1811.04968},
    
    publisher = {arXiv},
    author = {Bergholm, Ville and Izaac, Josh and Schuld, Maria and Gogolin, Christian and Ahmed, Shahnawaz and Ajith, Vishnu and Sohaib Alam, M. and Alonso-Linaje, Guillermo and AkashNarayanan, B. and Asadi, Ali and Arrazola, Juan Miguel and Azad, Utkarsh and Banning, Sam and Blank, Carsten and Bromley, Thomas R and Cordier, Benjamin A. and Ceroni, Jack and Delgado, Alain and Di Matteo, Olivia and Dusko, Amintor and Garg, Tanya and Guala, Diego and Hayes, Anthony and Hill, Ryan and Ijaz, Aroosa and Isacsson, Theodor and Ittah, David and Jahangiri, Soran and Jain, Prateek and Jiang, Edward and Khandelwal, Ankit and Kottmann, Korbinian and Lang, Robert A. and Lee, Christina and Loke, Thomas and Lowe, Angus and McKiernan, Keri and Meyer, Johannes Jakob and Montañez-Barrera, J. A. and Moyard, Romain and Niu, Zeyue and O'Riordan, Lee James and Oud, Steven and Panigrahi, Ashish and Park, Chae-Yeun and Polatajko, Daniel and Quesada, Nicolás and Roberts, Chase and Sá, Nahum and Schoch, Isidor and Shi, Borun and Shu, Shuli and Sim, Sukin and Singh, Arshpreet and Strandberg, Ingrid and Soni, Jay and Száva, Antal and Thabet, Slimane and Vargas-Hernández, Rodrigo A. and Vincent, Trevor and Vitucci, Nicola and Weber, Maurice and Wierichs, David and Wiersema, Roeland and Willmann, Moritz and Wong, Vincent and Zhang, Shaoming and Killoran, Nathan},
    month = nov,
    year = {2018},
    note = {ADS Bibcode: 2018arXiv181104968B},
}

@inproceedings{bartschi_grover_2020,
    title = {Grover {Mixers} for {QAOA}: {Shifting} {Complexity} from {Mixer} {Design} to {State} {Preparation}},
    shorttitle = {Grover {Mixers} for {QAOA}},
    url = {https://ieeexplore.ieee.org/abstract/document/9259965},
    doi = {10.1109/QCE49297.2020.00020},
    abstract = {We propose GM-QAOA, a variation of the Quantum Alternating Operator Ansatz (QAOA) that uses Grover-like selective phase shift mixing operators. GM-QAOA works on any NP optimization problem for which it is possible to efficiently prepare an equal superposition of all feasible solutions; it is designed to perform particularly well for constraint optimization problems, where not all possible variable assignments are feasible solutions. GM-QAOA has the following features: (i) It is not susceptible to Hamiltonian Simulation error (such as Trotterization errors) as its operators can be implemented exactly using standard gate sets and (ii) Solutions with the same objective value are always sampled with the same amplitude. We illustrate the potential of GM-QAOA on several optimization problem classes: for permutation-based optimization problems such as the Traveling Salesperson Problem, we present an efficient algorithm to prepare a superposition of all possible permutations of n numbers, defined on O (n 2) qubits; for the hard constraint k-Vertex-Cover problem, and for an application to Discrete Portfolio Rebalancing, we show that GM-QAOA outperforms existing QAOA approaches.},
    _urldate = {2026-07-16},
    booktitle = {2020 {IEEE} {International} {Conference} on {Quantum} {Computing} and {Engineering} ({QCE})},
    author = {Bärtschi, Andreas and Eidenbenz, Stephan},
    month = oct,
    year = {2020},
    keywords = {Approximation algorithms, Constraint optimization, Logic gates, Mixers, Optimization, Particle separators, Standards},
    pages = {72--82}
}

@article{preskill2018quantum,
  author    = {Preskill, John},
  title     = {Quantum {C}omputing in the {NISQ} era and beyond},
  journal   = {Quantum},
  volume    = {2},
  pages     = {79},
  year      = {2018},
  doi       = {10.22331/q-2018-08-06-79}
}

@article{cerezo2021variational,
  author    = {Cerezo, Marco and Arrasmith, Andrew and Babbush, Ryan and Benjamin, Simon C. and Endo, Suguru and Fujii, Keisuke and McClean, Jarrod R. and Mitarai, Kosuke and Yuan, Xiao and Cincio, Lukasz and Coles, Patrick J.},
  title     = {Variational quantum algorithms},
  journal   = {Nature Reviews Physics},
  volume    = {3},
  number    = {9},
  pages     = {625--644},
  year      = {2021},
  doi       = {10.1038/s42254-021-00348-9}
}

@article{qcomplexities,
  author        = {Bernstein, Ethan and Vazirani, Umesh},
  title         = {Quantum Complexity Theory},
  journal       = {SIAM Journal on Computing},
  volume        = {26},
  number        = {5},
  pages         = {1411--1473},
  year          = {1997},
  doi           = {10.1137/S0097539796300921},
  url           = {https://doi.org/10.1137/S0097539796300921},
  eprint        = {https://doi.org/10.1137/S0097539796300921}
}

@inbook{Karp1972,
  author        = "Karp, Richard M.",
  editor        = "Miller, Raymond E.
and Thatcher, James W.
and Bohlinger, Jean D.",
  title         = "Reducibility among Combinatorial Problems",
  booktitle     = "Complexity of Computer Computations: Proceedings of a symposium on the Complexity of Computer Computations, held March 20--22, 1972, at the IBM Thomas J. Watson Research Center, Yorktown Heights, New York, and sponsored by the Office of Naval Research, Mathematics Program, IBM World Trade Corporation, and the IBM Research Mathematical Sciences Department",
  year          = "1972",
  publisher     = "Springer US",
  address       = "Boston, MA",
  pages         = "85--103",
  isbn          = "978-1-4684-2001-2",
  doi           = "10.1007/978-1-4684-2001-2\_9",
  url           = "https://doi.org/10.1007/978-1-4684-2001-2\_9"
}

@book{schuld_machine_2021,
	address = {Cham},
	series = {Quantum {Science} and {Technology}},
	title = {Machine {Learning} with {Quantum} {Computers}},
	copyright = {https://www.springer.com/tdm},
	isbn = {978-3-030-83097-7 978-3-030-83098-4},
	url = {https://link.springer.com/10.1007/978-3-030-83098-4},
	doi = {10.1007/978-3-030-83098-4},
	language = {en},
	publisher = {Springer International Publishing},
	author = {Schuld, Maria and Petruccione, Francesco},
	year = {2021},
}

@article{mcclean_barren_2018,
	title = {Barren plateaus in quantum neural network training landscapes},
	volume = {9},
	issn = {2041-1723},
	url = {http://arxiv.org/abs/1803.11173},
	doi = {10.1038/s41467-018-07090-4},
	abstract = {Many experimental proposals for noisy intermediate scale quantum devices involve training a parameterized quantum circuit with a classical optimization loop. Such hybrid quantum-classical algorithms are popular for applications in quantum simulation, optimization, and machine learning. Due to its simplicity and hardware efficiency, random circuits are often proposed as initial guesses for exploring the space of quantum states. We show that the exponential dimension of Hilbert space and the gradient estimation complexity make this choice unsuitable for hybrid quantum-classical algorithms run on more than a few qubits. Specifically, we show that for a wide class of reasonable parameterized quantum circuits, the probability that the gradient along any reasonable direction is non-zero to some fixed precision is exponentially small as a function of the number of qubits. We argue that this is related to the 2-design characteristic of random circuits, and that solutions to this problem must be studied.},
	number = {1},
	_urldate = {2026-08-30},
	journal = {Nature Communications},
	author = {McClean, Jarrod R. and Boixo, Sergio and Smelyanskiy, Vadim N. and Babbush, Ryan and Neven, Hartmut},
	month = nov,
	year = {2018},
	note = {arXiv:1803.11173 [quant-ph]},
	keywords = {Computer Science - Machine Learning, Quantum Physics, Physics - Chemical Physics},
	pages = {4812},
	file = {Preprint PDF:/home/manu/Zotero/storage/B2XMVP8H/McClean et al. - 2018 - Barren plateaus in quantum neural network training landscapes.pdf:application/pdf;Snapshot:/home/manu/Zotero/storage/96BAZHZC/1803.html:text/html},
}

@article{egger_warm-starting_2021,
  author    = {Egger, Daniel J. and Mare{\v{c}}ek, Jakub and Woerner, Stefan},
  title     = {Warm-starting quantum optimization},
  journal   = {Quantum},
  volume    = {5},
  pages     = {479},
  year      = {2021},
  doi       = {10.22331/q-2021-06-17-479}
}

@article{brandao2018foraging,
  title={Foraging for quantum advantage},
  author={Brandao, Fernando G. S. L. and Crosson, Michael and Mbeng, Glen Bigan and Schaeffer, Tameem},
  journal={arXiv preprint arXiv:1812.04170},
  year={2018}
}

@article{rajak_quantum_2023,
  title = {Quantum {Annealing}: {An} {Overview}},
  volume = {381},
  issn = {1364-503X, 1471-2962},
  shorttitle = {Quantum {Annealing}},
  url = {http://arxiv.org/abs/2207.01827},
  doi = {10.1098/rsta.2021.0417},
  abstract = {In this review, after providing the basic physical concept behind quantum annealing (or adiabatic quantum computation), we present an overview of some recent theoretical as well as experimental developments pointing to the issues which are still debated. With a brief discussion on the fundamental ideas of continuous and discontinuous quantum phase transitions, we discuss the Kibble-Zurek scaling of defect generation following a ramping of a quantum many body system across a quantum critical point. In the process, we discuss associated models, both pure and disordered, and shed light on implementations and some recent applications of the quantum annealing protocols. Furthermore, we discuss the effect of environmental coupling on quantum annealing. Some possible ways to speed up the annealing protocol in closed systems are elaborated upon: We especially focus on the recipes to avoid discontinuous quantum phase transitions occurring in some models where energy gaps vanish exponentially with the system size.},
  number = {2241},
  _urldate = {2026-09-01},
  journal = {Philosophical Transactions of the Royal Society A: Mathematical, Physical and Engineering Sciences},
  author = {Rajak, Atanu and Suzuki, Sei and Dutta, Amit and Chakrabarti, Bikas K.},
  month = jan,
  year = {2023},
  note = {arXiv:2207.01827 [cond-mat.stat-mech]},
  keywords = {Condensed Matter - Statistical Mechanics, Quantum Physics},
  pages = {20210417},
}

@article{sack2021quantum,
  title={Quantum annealing initialization of the quantum approximate optimization algorithm},
  author={Sack, Stefan H. and Serbyn, Maksym},
  journal={Quantum},
  volume={5},
  pages={491},
  year={2021},
  publisher={IOP Publishing}
}

@article{zhou2020quantum,
  title={Quantum approximate optimization algorithm: Performance, parameter schedules, and metric structures},
  author={Zhou, Leo and Wang, Sheng-Tao and Choi, Soonwon and Pichler, Hannes and Lukin, Mikhail D.},
  journal={Physical Review X},
  volume={10},
  number={2},
  pages={021067},
  year={2020},
  publisher={APS}
}
\newpage
\appendix

\section{Quantum Approximate Optimization Algorithm}\label{sec:qaoa}
As research into \gls{qc} advances, more application areas are becoming viable in the search for quantum advantage, opening the door to combining \gls{qc} and \gls{ml} in what is formally known as \gls{qml}~\citep{schuld_machine_2021}. \gls{qml} applies concepts from \gls{ml} to the computational model provided by quantum computers, enabling novel algorithms. A prime example is \gls{qaoa}, an algorithm similar to \gls{qa} and suitable for gate-based computing. In \gls{qa}, the system is initialized in the known ground state of a Hamiltonian $H_M$; then the controlled evolution $H(t) = \frac{t}{T} H_C + \left(1 - \frac{t}{T}\right)H_M$ is applied, gradually evolving the system into the one defined by the cost Hamiltonian $H_C$. According to the adiabatic theorem, if the process is slow enough (i.e., $T$ is large), the system remains in its ground state throughout the evolution. If the cost Hamiltonian is constructed so that its ground state corresponds to the desired target, then the solution can be retrieved after the evolution.

While adiabatic computing can retrieve the ground state of an arbitrary Hamiltonian, the time $T$ required to satisfy the adiabatic theorem may render the algorithm unsuitable for certain tasks. A different approach arises by mimicking this evolution using \gls{qaoa}: the continuous evolution of the system is discretized using the Trotter–Suzuki decomposition~\citep{farhi_quantum_2014}, which decomposes the evolution into several discrete steps:
\begin{equation}\label{eq:trotterization}
    U(T) \approx \prod_{k=1}^P e^{-i\Delta t\left(1-\frac{t_k}{T}\right)H_M}\cdot e^{-i\Delta t\frac{t_k}{T}H_C},
\end{equation}
in which $P$ denotes the depth and accuracy of the approximation, $\Delta t$ is the timestep size $\Delta = \frac{T}{P}$, and $t_k$ are the various discrete time steps. However, using a $P$ large enough to cause a good approximation is prohibitive, so \gls{qaoa} replaces the pre-scheduled time steps with learnable parameters, resulting in \begin{equation}
    U(T) \approx U(\gamma, \beta) = \prod_{k=1}^P e^{-i\beta_kH_M}\cdot e^{-i\gamma_kH_C}. 
\end{equation}
The main components of this equation are the two Hamiltonian terms $H_M$ and $H_C$, called respectively the \textit{mixer} and \textit{cost} Hamiltonian. The former represents our initial system with a known ground state, while the latter encodes the \gls{qubo} problem 
\begin{equation}
    \min_x x^TQx+g^Tx    
\end{equation}
into the Ising model given by \begin{equation}
    H_C = \sum_{i,j}J_{ij}Z_iZ_j + \sum_i h_i Z_i.
\end{equation}
The interaction coefficients can be retrieved using the following relation 
    \begin{gather}
        J_{ij} = -\frac{Q_{ij}}{4}, \\
        h_i = \frac{g_i}{2} + \sum_j\frac{Q_{ij}+Q_{ji}}{4}.
    \end{gather}
The mixer Hamiltonian is usually defined as $H_M=-\sum_iX_i$, which has the lowest energy state $\ket{+}$. Even though other proposals exist~\citep{wang_xy_2020, bartschi_grover_2020}, for the rest of the paper, 
this results in a variational circuit defined as follows:
\begin{equation}
\begin{aligned}
    U(\gamma, \beta) = \prod_{k=1}^P \Biggr(&\prod_iR_{X_i}(-\beta_k)\prod_iR_{Z_i}(\gamma_kh_i) \\
    &\prod_{ij}R_{Z_iZ_j}(\gamma_k J_{ij})\Biggr).
\end{aligned}
\end{equation}
By using this discrete evolution, the goal becomes minimizing the expectation value of the cost Hamiltonian:
\begin{equation}
    \min_{\gamma, \beta} f(\gamma, \beta) =\bra{\psi(\gamma, \beta)}H_C\ket{\psi(\gamma, \beta)},
\end{equation}
with $\ket{\psi(\gamma,\beta)}=U(\gamma,\beta)\ket{+}$.
This can be achieved using classical optimization algorithms based on \gls{gd}, updating the parameters using the gradient of the cost function:
\begin{equation}
    \theta_{t+1} = \theta_t - \eta_t\nabla_{\theta_t} f(\gamma, \beta),
\end{equation}
where $\theta_t$ represents the set of parameters $(\gamma, \beta)$ at the $t$-th optimization step. In this setting, the derivatives can be computed using the parameter-shift rule~\citep{wierichs_general_2022}, which allows exact computation of partial derivatives using multiple circuit evaluations:
\begin{equation}
    \frac{\partial f(\theta)}{\partial\theta^{(i)}} = \frac{f\!\left(\theta + \epsilon_i \frac{\pi}{2}\right)-f\!\left(\theta - \epsilon_i \frac{\pi}{2}\right)}{2},
\end{equation}
with $\epsilon_i$ representing the vector with all $0$ entries except for the $i$-th component, which is set to $1$.

The optimization procedure continues until convergence of the cost function, i.e., until the lowest-energy state is found. Then, by sampling the most probable states, we can retrieve the solution to the problem.

All our training procedures use the \gls{gd}-based optimizer Adam~\citep{kingma_adam_2017}, and the circuit is trained for an indefinite number of steps, with the exit condition being that the Ising energy has not been reduced by at least $10^{-2}$ in the last $20$ steps. If not stated otherwise, the simulations are performed using the \texttt{lightning.qubit} statevector simulator provided by PennyLane~\citep{bergholm_pennylane_2018}.

\end{document}